\documentclass[apjl]{emulateapj}
\usepackage{apjfonts}

\shorttitle{Planetary Mass Candidate Companion to a Young Star}
\shortauthors{Lafreni\`{e}re et al.}

\newcommand\planethost{1RXS~J160929.1-210524}
\newcommand\primary{1RXSJ1609-2105}

\begin{document}

\title{Direct Imaging and Spectroscopy of a Planetary Mass Candidate Companion to a Young Solar Analog}

\author{David Lafreni\`ere, Ray Jayawardhana and Marten H. van Kerkwijk}
\affil{Department of Astronomy and Astrophysics, University of Toronto, 50 St. George Street, Toronto, ON, M5S 3H4, Canada}
\email{lafreniere@astro.utoronto.ca}

\begin{abstract}
We present Gemini near-infrared adaptive optics imaging and spectroscopy of a planetary mass candidate companion to \planethost, a roughly solar-mass member of the 5~Myr-old Upper Scorpius association. The object, separated by 2.22\arcsec\ or 330~AU at $\sim$150~pc, has infrared colors and spectra suggesting a temperature of $1800_{-100}^{+200}$~K, and spectral type of L4$_{-2}^{+1}$. The $H$- and $K$-band spectra provide clear evidence of low surface gravity, and thus youth. Based on the widely used DUSTY models, we infer a mass of $8_{-2}^{+4}~M_{\rm Jupiter}$. If gravitationally bound, this would be the lowest mass companion imaged around a normal star thus far, and its existence at such a large separation would pose a serious challenge to theories of star and planet formation.
\end{abstract} 
\keywords{stars: pre--main sequence --- stars: low-mass, brown dwarfs --- planetary systems}

\section{Introduction}

Since 1995, over 300 planets have been identified around stars other than the Sun,\footnote{See {\it The Extrasolar Planets Encyclopaedia} at \url{http://exoplanet.eu/}} revealing a remarkable diversity in the properties of planets as well as the architecture of planetary systems. Yet, the three hitherto successful planet detection techniques -- radial velocity monitoring, transit searches and microlensing surveys -- are generally limited to planets in orbits much smaller than the full radial extent of our solar system. Thus, the discoveries to date, while dramatic, can only provide an incomplete picture of the overall planet population. Direct imaging can help complete the census by uncovering planets at wide separations. Furthermore, it enables extensive follow-up studies to characterize the planets. However, due to the small angular separation and the high brightness contrast between a planet and its star, direct imaging has not succeeded thus far \citep[e.g.][and references therein]{lafreniere07}. At present, our best hope lies in targeting nearby young stars for newborn planets.

During their formation, giant planets and brown dwarfs (BDs), like stars, generate heat from gravitational contraction. But, unlike stars, which eventually reach core temperatures sufficient for hydrogen fusion, these objects are left without a means of producing energy, and thus rapidly cool down and become dimmer with time. Therefore younger planets and BDs, as companions to stars, are much easier to detect directly than their older counterparts. Accordingly, the lowest mass companions imaged so far all orbit stars younger than $\sim$30~Myr; several of them (e.g. CHXR~73~B, \citealp{luhman06}; AB~Pic~B, \citealp{chauvin05b}; DH~Tau~B, \citealp{itoh05}; GQ~Lub~B, \citealp{neuhauser05}) have masses slightly above the deuterium burning limit of 13~$M_{\rm Jupiter}$, which is often used as a boundary to differentiate planets from the more massive BDs\footnote{{\it Working Group on Extrasolar Planets of the International Astronomical Union}, see \url{http://www.dtm.ciw.edu/boss/definition.html}}, though some researchers prefer a distinction based on the formation mechanism, whereby planets form within a circumstellar disk and BDs form through cloud fragmentation. One lower mass companion ($\sim$8~$M_{\rm Jupiter}$) orbits the young 25~$M_{\rm Jupiter}$ BD 2MASSW~J1207334-393254, rather than a star \citep{chauvin04, mohanty07}; other similar systems might exist \citep[e.g.][]{bejar08}. All of these low-mass sub-stellar companions are located at large separations from their primaries, in sharp contrast with solar system planets and extra-solar planets detected by indirect methods, possibly reflecting a fundamental difference in their formation mechanism.

Here we report the direct imaging discovery of a 6--12~$M_{\rm Jupiter}$ candidate companion to a young solar-mass star, \planethost, in the nearby ($145\pm20$~pc), 5~Myr-old Upper Scorpius association \citep{preibisch02}.

\section{Observations and data reduction}\label{sect:obs}

\subsection{Imaging}

Initial imaging of \planethost\ was done on 2008 April 27 in $K_{\rm s}$ using the NIRI camera \citep{hodapp03} and the ALTAIR adaptive optics system \citep{herriot00} at the Gemini North Telescope.  We used 5 dither positions, with at each position one co-addition of twenty 0.3~s integrations in fast, high read-noise mode, and one single 10~s integration in slow, low read-noise mode, thus at each position providing an unsaturated image of the target star and a much deeper image of the field that can be registered and scaled in flux without ambiguity.  Follow-up imaging in $H$ and $J$ was obtained on 2008 June 21 using the same instrument.  For $H$, we again used 5 dither positions with thirty 0.2~s and a single 10~s co-addition at each position, and for $J$ we used 7 dither positions with ten 0.5~s and a single 10~s co-addition.

The imaging data were reduced using custom {\em IDL} routines.  We subtracted a sky frame constructed using the median of the images at all dither positions (with regions dominated by the target's signal masked), divided by a normalized flat-field, and replaced bad pixels by a median over their neighbors.  Next, we merged the long- and short-exposure images, properly scaled in intensity, and co-aligned and co-added the resulting images.  A composite color image is shown in Fig.~\ref{fig1}.

\subsection{Spectroscopy}

Spectroscopic follow-up was done on 2008 June 21, using NIRI in grism mode with Altair at the Gemini North Telescope.  We used the $f/32$ 10-pixel wide (0.214\arcsec) slit, rotated to obtain spectra of the primary and companion simultaneously.  We obtained five exposures of 180~s in $K$ and 210~s in $H$, at nods separated by 4.44\arcsec\ along the slit.  For telluric and instrumental transmission correction, the A0V star HD~138813 was observed in $K$ before the science target, and HD~151787 in $H$ afterwards.  To confirm the reliability of the data, further sets of $H$ and $K$ spectra were taken on 2008 August 21 and 2008 August 24, respectively.  This time, in $K$, twelve 180~s exposures were taken in four groups of three 4.44\arcsec\ nods, each group being further nodded by 0.3\arcsec, and in $H$, nine 180~s exposures were obtained in three similarly nodded groups.  For calibration, HD~151787 was observed after the science target for both sequences.

With the wide slit, we minimize effects from the chromatic adaptive-optics-corrected point-spread function, atmospheric differential refraction ($<0.03\arcsec$ in-band), or small errors in nod position.  Based on our imaging, slit losses should vary by at most 5\% between $H$ and $K$.  Thus, even if the standards are taken with slightly different image quality, residual slit effects in the calibrated fluxes should be small.  With the wide slit, the point-spread function sets the resolution of $\sim\!12$~\AA\ in $H$ and $\sim\!18$~\AA\ in $K$.

The data were reduced using custom {\em IDL} routines.  We first subtracted the sky background, determined from exposures at different nods, divided by a normalized flat field, and masked bad pixels. The images were then rectified, using cubic interpolation, for the slight curvature of the traces.  We extracted optimally-weighted fluxes using the normalized trace of the spectrum, constructed separately for each image and allowed to vary slowly with wavelength.  We chose a 2.0\arcsec\ width for the trace, for which chromatic effects should be less than a few percent.  The companion's spectrum was extracted using the same trace, but shifted in position, truncated to 0.25\arcsec, and properly re-normalized to avoid introducing additional chromatic effects. Prior to extraction, any remaining flux from the primary was removed by subtracting a straight line along the spatial direction fitted to both sides of the companion's trace.  Wavelength calibration was done using exposures of an Ar arc lamp.  Next, we divided the spectra by those of the telluric standard, corrected for their spectral slope using a 9520~K blackbody curve, and with hydrogen absorption lines removed by dividing by Voigt profile fits to each line.  Since the $K$- and $H$-band spectra obtained in June and August were very similar, we co-added them to improve the signal-to-noise ratio.  Finally, we flux-calibrated the spectra relative to the 2MASS magnitudes of the primary (using the spectral response and zero points given in \citealt{cohen03}).  Synthetic 2MASS contrasts and colors for the companion computed from the flux-calibrated spectra agree within a few percent with the values obtained from the photometry.

\section{Analysis and results}\label{sect:analysis}

The position of the companion relative to the primary was found by fitting a 2-D Gaussian model to both the primary and companion PSFs; the orientation of the image was obtained from the FITS header. The measurements uncertainties were estimated from the dispersions of the measurements made on all the individual images. We obtained a separation of $103.70\pm0.06$ pixels, corresponding to $2.219\arcsec\pm0.002\arcsec$ given the pixel scale of 0.0214\arcsec, and a position angle of $27.7\pm0.1$~deg. Systematic uncertainties are likely larger than measurement uncertainties; based on previous experience with similar observations we estimate them at $\sim$0.03\arcsec\ for the separation and $\sim$0.5~deg for the position angle. 

\begin{figure}
\epsscale{1}
\plotone{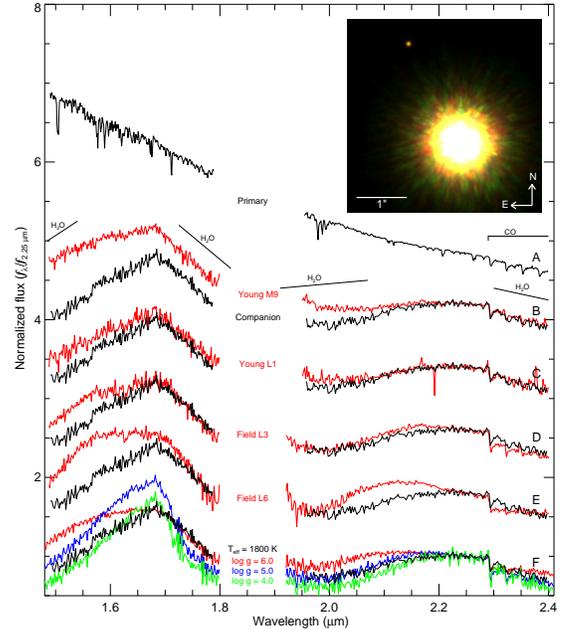}
\caption{\label{fig1} Spectra of \planethost\ and its faint candidate companion.  The primary's spectrum (row A) is as expected for a K7 spectral type. The candidate companion's spectrum (black curves repeated in rows B--F) is compared with the spectra of two young BDs (red curves on rows B--C; M9, USco~J160830-233511; and L1, USco~J163919-253409) and two older, cooler field BDs (red curves on rows D--E; L3, 2MASSW~J1506544+132106; and L6, 2MASSW~J1515008+484742), as well as with theoretical spectra with different surface gravities (colored curves in row F). The spectra in rows B--F are binned to a resolving power of $\sim$850 and normalized at 2.25~$\mu$m. The inset at the top right shows our composite image of the two objects. Blue, green, and red represent images taken in $J$, $H$, and $K_{\rm s}$, with intensities scaled such that they are proportional to the photon rates inferred from the 2MASS magnitudes of the primary.}
\end{figure}

The relative photometry was computed using aperture photometry with a radius of one PSF FWHM. An azimuthally symmetric median intensity profile was subtracted from the images prior to measuring the flux of the companion to avoid contamination from the primary. The uncertainties were estimated from the dispersion of the measurements made on all of the individual images in each filter. The results are shown in Table~1; the near-infrared colors of the companion suggest a mid-L spectral type. For completeness, we note that four other faint sources were detected farther from the primary, but these have $J-K_{\rm s}<1.1$ and thus are likely background stars.

The spectra of both the primary and its candidate companion are shown in Fig.~\ref{fig1}, alongside template spectra of known BDs in Upper Scorpius\footnote{see http://www.iac.es/galeria/nlodieu/publications.html} \citep{lodieu08}, field dwarfs\footnote{See http://irtfweb.ifa.hawaii.edu/$\sim$spex/spexlibrary/IRTFlibrary.html} \citep{cushing05} and synthetic spectra from the DUSTY models \citep{chabrier00}. The spectrum of the companion confirms that it is very cool, showing important water vapor absorption on either sides of the $H$ and $K$ bands and strong CO band heads beyond 2.29~$\mu$m. Compared with field dwarfs, the $H$-band spectrum of our companion has a much more triangular shape which is likely caused by lower surface gravity, as evidenced by the much better agreement with the young Upper Scorpius BDs and the low-gravity model spectra. Indeed, the model spectra show that the blue side of the $H$-band spectrum is suppressed as the surface gravity decreases; a similar effect is present at the blue side of the $K$ band. Also, the triangular $H$-band profile is a well-known indicator of low surface gravity that has been observed and discussed in many instances in the literature \citep[e.g.][]{martin03,kirkpatrick06}. Lower surface gravity would be expected for an object that has not yet fully contracted, and thus these features confirm the youth of the candidate companion. Uncertainties in the model opacities could account for the mismatch on the red side of the $H$-band for low gravities, and thus this should not be taken as evidence of high gravity \citep{leggett01}. Overall, reasonable fits of DUSTY model spectra are obtained with effective temperatures of 1700-1900~K and $\log{g}$ of 3.5--5.0, with a best fit for 1800~K. 

The $J-K_{\rm s}$ color of the companion complements the $HK$ spectroscopy and offers one more check of its temperature. Although low-gravity BDs generally appear redder than their older counterparts for a given spectral type \citep{kirkpatrick08}, all but one of the known low-gravity BDs earlier than L2 in the sample of \citet{kirkpatrick08} are bluer than our companion; the exception being the extremely peculiar 2MASS~J01415823-4633574 (L0pec, $J-K_{\rm s}=1.74$). Similarly, all Upper Scorpius BDs (M8--L2) in the sample of \citet{lodieu08}, except for two L1, have bluer colors. Finally, most BDs later than L5, particularly those with lower gravity, have redder colors. Thus the $J-K_{\rm s}$ color indicates a spectral type of L2--L5. 

\begin{figure}
\epsscale{1}
\plotone{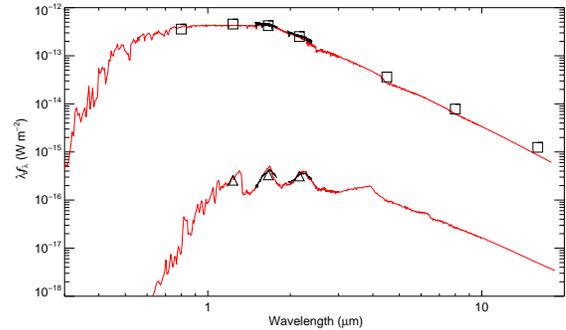}
\caption{\label{fig2} Spectral energy distribution of \planethost\ (top) and its faint candidate companion (bottom).  The black curves are our spectra, the squares photometry of the primary taken from the literature, and the triangles photometry of the companion inferred from our magnitude differences (see Table~\ref{tbl1}). Overdrawn in red are models.  For the primary, we used a NextGen model spectrum with $T_{\rm eff}=4000$~K and $\log{g}=4.0$, and a radius of $1.352\,R_\odot$; for the companion, we used a DUSTY model spectrum with $T_{\rm eff}=1800$~K and $\log{g}=4.0$, and a radius of $0.171\,R_\odot$. For both, we scaled the model fluxes to a distance of 150~pc. The slight excess at long wavelengths seen for the primary might indicate the presence of a small residual disk, although no evidence is seen for ongoing accretion (see text).  Alternatively, it may simply reflect uncertainties in the model, our inferred temperature, or the reddening.}
\end{figure}

Given the above considerations, and based primarily on the comparison with model spectra, we infer $T_{\rm eff}=1800_{-100}^{+200}$~K, corresponding to a spectral type of L4$_{-2}^{+1}$ given the $T_{\rm eff}$--spectral type relation of \citet{golimowski04}. The age of the Upper Scorpius association is well constrained at 5~Myr with no significant age spread: all star formation seemingly occurred over a period of less than 1~Myr \citep{preibisch02}. This value of 5~Myr is consistently obtained from isochrone fitting using several different evolution models, and is in good agreement with the $\sim$4.5~Myr kinematic age of the association, obtained by tracing back the proper motions of massive stars \citep{blaauw78}. Given an age estimate of $5\pm1$~Myr and $T_{\rm eff}=1800_{-100}^{+200}$~K for the candidate companion, the DUSTY evolutionary models yield a mass of $0.008_{-0.002}^{+0.004}~M_\odot$, below the deuterium burning limit. The mass uncertainty quoted reflects the widest interval obtained for the quoted ranges of age and $T_{\rm eff}$. As an additional test, we can compare not just the temperature (colors), but also the magnitudes (luminosity) of the object against the model predictions: the best agreement [$(J,H,K)_{\rm obs}-(J,H,K)_{\rm model}=(0.10,0.02,0.06)$] is obtained for a mass of 0.008~$M_\odot$, an age of 5~Myr and a distance of 150~pc (see Fig.~\ref{fig2}). Factoring in the uncertainties on the photometry measurements as well as those on the age and distance, this approach yields a mass estimate of 0.007--0.012~$M_\odot$. As yet another check, we consider the evolution models of \citet{burrows97}; for the above $T_{\rm eff}$ and age estimates, these models yield a mass of 0.007--0.0011~$M_\odot$. Of course, despite this, some fundamental shortcomings, or lack of absolute calibration, in the models used for the estimates of age, temperature and mass could result in higher uncertainties than those presented above, but given the difficulty in quantifying such errors we can only alert the readers to this possibility.

The primary star was classified as a member of the Upper Scorpius association by \citet{preibisch98} and \citet{preibisch99} based on lithium absorption, X-ray emission, and position in the Hertzsprung-Russell diagram; these authors determined a spectral type of M0V based on low-resolution optical spectroscopy. New high-resolution optical spectra suggest a slightly earlier spectral type, K7V$\pm1$, based on a comparison of several atomic and molecular lines with spectral templates (D.C. Nguyen, private communication). This is also consistent with our near-infrared spectrum, and corresponds to $4060_{-200}^{+300}$~K \citep{sherry04}. Furthermore, for this effective temperature, the predictions of the NextGen evolutionary models \citep{baraffe98,baraffe02}, using $\alpha_{\rm mix}=1.9$, are in very good agreement with the photometric measurements of the primary for an age of 5~Myr and a distance of 150~pc (see Fig.~\ref{fig2}), and indicate a mass of $0.85_{-0.10}^{+0.20}$~$M_\odot$. We found no evidence for on-going accretion in the optical and near-infrared spectra, nor signs of significant reddening, consistent with the value of $A_V=0$ reported by \citet{preibisch99}. \citet{carpenter06} reports no infrared excess at 8.0~$\mu$m and 16.0~$\mu$m (see also Fig.~\ref{fig2}) based on {\em Spitzer} observations; this indicates that there is little if any disk material left around the star, consistent with the absence of any spectroscopic signatures of on-going mass accretion. Our adaptive optics images indicate that the primary is not itself a near-equal luminosity binary with a separation larger than $\sim\!0.06$\arcsec\ ($>\!9$~AU). A summary of the properties of the primary and companion is given in Table~\ref{tbl1}.

\begin{deluxetable}{lcc}
\tablewidth{0pt}
\tablecolumns{3}
\tablecaption{\label{tbl1} Properties of \planethost\ A{\rm b}}
\tablehead{
\colhead{} & \multicolumn{2}{c}{Value} \\
\cline{2-3}
\colhead{Parameter} & \colhead{Primary} & \colhead{Companion}
}
\startdata
Angular separation (\arcsec) & \multicolumn{2}{c}{$2.219\pm0.002$} \\
Position angle (deg) & \multicolumn{2}{c}{$27.7\pm0.1$} \\
$\Delta J$ (mag)         & \multicolumn{2}{c}{$8.08\pm0.12$} \\
$\Delta H$ (mag)         & \multicolumn{2}{c}{$7.75\pm0.07$} \\
$\Delta K_{\rm s}$ (mag) & \multicolumn{2}{c}{$7.25\pm0.18$} \\
$I$ (mag)\tablenotemark{a}        & $10.99\pm0.03$ & \nodata \\
$J$ (mag)\tablenotemark{b}        & $9.820\pm0.027$ & $17.90\pm0.12$ \\
$H$ (mag)\tablenotemark{b}        & $9.121\pm0.023$ & $16.87\pm0.07$ \\
$K_{\rm s}$ (mag)\tablenotemark{b} & $8.916\pm0.021$ & $16.17\pm0.18$ \\
$S_\nu$ [4.5$\mu$m] (mJy)\tablenotemark{c} & $54.4\pm0.7$ & \nodata \\
$S_\nu$ [8.0$\mu$m](mJy)\tablenotemark{c} & $20.8\pm0.1$ & \nodata \\
$S_\nu$ [16.0$\mu$m] (mJy)\tablenotemark{c} & $6.66\pm0.04$ & \nodata \\
$J-K_{\rm s}$ (mag) & $0.830\pm0.034$ & $1.73\pm0.22$ \\
$H-K_{\rm s}$ (mag) & $0.205\pm0.031$ & $0.70\pm0.19$ \\
Spectral type & K7V$\pm$1 & L4$_{-2}^{+1}$ \\
$T_{\rm eff}$ (K) & $4060_{-200}^{+300}$\tablenotemark{d} & $1800_{-100}^{+200}$\tablenotemark{e} \\
Radius ($R_\odot$)\tablenotemark{e} & $1.352$ & $0.171$ \\
Mass ($M_\odot$)\tablenotemark{e} & $0.85_{-0.10}^{+0.20}$ & $0.008_{-0.002}^{+0.004}$ \\
Projected separation (AU)\tablenotemark{e} & \multicolumn{2}{c}{$\sim$330}
\enddata
\tablenotetext{a}{From the DENIS catalog.}
\tablenotetext{b}{From the 2MASS PSC and our contrast measurements.}
\tablenotetext{c}{From \citet{carpenter06}.}
\tablenotetext{d}{From the spectral types based on \citet{sherry04}.}
\tablenotetext{e}{From the models of \citet{baraffe98,baraffe02} for the primary and \citet{chabrier00} for the companion.}
\tablenotetext{f}{Assuming a distance of 150~pc \citep{preibisch02}.}
\end{deluxetable}

\section{Discussion}\label{sect:discussion}

Although our photometry and spectroscopy establish that the candidate companion has low gravity and a mass below 0.012~$M_\odot$, they do not prove that it is physically bound to the primary star rather than a free-floating planet in the association. Based on integration of their best-fit mass function, \citet{preibisch02} estimate that the entire Upper Scorpius population comprises 2525 stars more massive than $>0.1$~$M_\odot$, distributed over an area of $\sim$150~deg$^2$. Assuming very conservatively that there are as many free-floating planets in the association as there are stars $>0.1$~$M_\odot$, the probability that one would fall within 2.5\arcsec\ from any of the 85 stars we have observed would be only 0.002. Thus this scenario is unlikely. Verification of common proper-motion over the next few years will nevertheless be important, although it would not readily confirm physical association of the two objects given the small internal velocity dispersion of the association. The latter would require detection of orbital motion, which could take several years given the small orbital motion expected ($\sim$2~mas~yr$^{-1}$).

Previous direct imaging surveys for planets around nearby solar-type stars have put upper limits of $\sim$6\% on the fraction of stars with planets more massive than 5~$M_{\rm Jupiter}$ at separations over $\sim$50~AU \citep[][and references therein]{lafreniere07}. Our single candidate in a sample of $\sim$85 stars is consistent with these results, and confirms that planets on wide separations are rare also at ages of a few Myr.

Compared to known star-planet systems, the inferred mass for the candidate companion of \planethost\ is near the high end of the range. What stands out even more is the large separation of $>$300~AU. The simplest explanations would be that there is no physical association or that the companion has recently been ejected. Both appear unlikely, however, and thus it seems worthwhile to speculate about possible formation scenarios. In situ growth via core accretion \citep{pollack96} appears unlikely: the formation timescale would greatly exceed the age of the system, even if a disk could survive long enough. On the other hand, at smaller separations this mechanism could build a $\sim$5~$M_{\rm Jupiter}$ planet in 1--5~Myr \citep{pollack96,alibert05}, which might then migrate outwards through interactions with a disk or other massive planets (either can be quite rapid; \citealt{peplinski08,veras04}). A problem with this alternative, however, is that a planet formed through core accretion may not be able to reach the observed temperature \citep{marley07}. In situ formation via gravitational instability \citep{boss97} is another possibility, but it would require a rather massive ($\gtrsim$0.2-0.35 M$_\star$) disk. Typical disks around T Tauri stars contain only 0.01--0.1 M$_\star$ of material \citep{scholz06}, but more massive disks may exist during earlier proto-stellar phases. Still, the thermal state of the disk may prevent fragmentation even at radii as large as required here \citep{matzner05}. Lastly, this system may have formed like a binary star, through the fragmentation of a pre-stellar core.  Current star formation simulations, however, find it difficult to make binaries with extreme mass ratios and very low mass components \citep{bate03}. 

Future observations could look for additional closer-in companions, evidence of a large debris disk, and whether the object we found is in a highly eccentric or nearly circular orbit. If bound, the very existence of the \primary\ system poses a challenge to theories of planet and star formation, and may well suggest that there is more than one mechanism in nature for producing planetary mass companions around normal stars.

\acknowledgments

We thank the Gemini staff, particularly Jean-Ren\'e Roy and Andrew Stephens, for help and support with the observations, Duy Nguyen, Kaitlin Kratter, Subhanjoy Mohanty, and Adam Kraus for discussions, and Mark Marley for pointing out possible problems with core accretion formation. This work is supported in part through an FQRNT fellowship to DL, and NSERC grants to RJ and MHvK. 


\end{document}